\documentstyle[12pt]{article}

\begin{document}

\pagestyle{empty}

\noindent {\small USC-97/HEP-B2\hfill \hfill hep-th/9704054}\newline
{\small \hfill CERN-TH/97-66}

{\vskip 1.5cm}

\begin{center}
{\LARGE A Case for 14 Dimensions\ } \\[0pt]

{\vskip 1cm}

{\bf Itzhak Bars}$^{a,b}$ {\Large \ \\[0pt]
}

{\vskip .6cm}

{\bf TH Division, CERN, CH-1211 Geneva 23, Switzerland}

{\vskip 1cm}

{\bf ABSTRACT}
\end{center}

\noindent
Extended superalgebras of types $A,B,C,$ heterotic and type-I are all
derived as solutions to a BPS equation in $14$ dimensions with signature $
\left( 11,3\right) .$ The BPS equation as well as the solutions are
covariant under SO$\left( 11,3\right) .$ This shows how supersymmetries with 
$N\leq 8$ in four dimensions have their origin in the same superalgebra in
14D. The solutions provide different bases for the same superalgebra in 4D,
and the transformations among bases correspond to various dualities.

\vfill
\hrule width 6.7cm \vskip 2mm

$^a$ {\small Research supported in part by the U.S. Department of Energy
under grant number DE-FG03-84ER40168.}

{$^b$ {\small On sabbatical leave from the Department of Physics and
Astronomy, University of Southern California, Los Angeles, CA 90089-0484,
USA.}}

\vfill\eject

\setcounter{page}1\pagestyle{plain}

\medskip

\section{\protect\medskip 64 supercharges}

It is well known that in four flat dimensions there cannot be more than
eight conserved real (Majorana) supercharges. If one imagines that the 4D
theory comes from some fundamental theory, then the fundamental theory
apparently may not have more than 32 {\it real} supercharges in its flat
limit: If it had more than 32 it would imply more than $N=8$ in 4D. The
physical basis for this assertion is that in flat 4D spacetime there cannot
be massless interacting particles with helicity higher than 2 and/or that
there is only one graviton. For $N\geq 8,$ a supermultiplet that includes
the graviton in 4D necessarily contradicts these facts.

A caveat in this argument is that there may be different sets of 32
supercharges that are equivalent to each other under duality symmetries from
the point of view of a lower dimensional effective theory. We suggest that
the existence of dualities may allow 64 supercharges, and that in dimensions
10 to14 one can embed three sets 32$_A,$32$_B$ and 32$_C$ as different
projections of the 64, which form three distinct superalgebras of types $
A,B,C$. First examples of theories containing the 32$_{A,B}$ supercharges
are 10D supergravity/superstrings of type-IIA,B. In 11D supergravity there
is 32$_A.$ For dimensions 9 and below the distinction between 32$_{A,B}$
disappears, but the T-duality which is well known for $d\leq 9$ actually
performs the transformation 32$_A$ $\longleftrightarrow $ 32$_B.$ This
duality was interpreted as a transformation of 64 spinors among themselves
by a transformation of hidden dimensions \cite{stheory}. Since the various
types of non-perturbative dualities map different forms of theories into
each other it is not clear whether the overall theory behind all of it is a
theory with 64 supercharges or a theory with only 32 of them.

Are there 64 supercharges rather 32? What mechanism generates effectively 32
supercharges in different sectors? As suggested below 64$\rightarrow 32$
happens naturally through a BPS condition that can be formulated as a
covariant equation in 14 dimensions and which has three distinct branches of
solutions labelled by $A,B,C.$ In this way we answer a related set of
questions: Is there a single set of 528 bosons in the superalgebras $A,B,C$
or are there three sets 528$_{A,B,C}$ that are mapped to each other by
dualities. If they are different, what is the subset of bosons that is
common to various sectors?

In this paper we explore further the possibility that the unknown
fundamental theory behind non-perturbative string theory and its dualities
may be a theory more usefully formulated in higher dimensions, perhaps in
14D. By recognizing its hidden dimensions one may better understand its
overall structure as well as its low energy properties. Since the
fundamental theory is unknown we concentrate only on properties of its
supersymmetries. We assume that in the fundamental theory 32$_A$ and 32$_B$
are two distinct sets mapped to each other by a transformation that is
interpreted as T-duality from the point of view of effective theories in
lower dimensions. Since T-duality maps small to large distances, our
assumption implies that some of the 32$_A$ supercharges that govern
supersymmetry at large distances get mapped to some of the 32$_B$
supercharges that govern supersymmetry at small distances, and vice versa.
In this paper we show how to embed type $A,B,C,$ heterotic and type-I
superalgebras {\it covariantly} in the framework of 14 dimensions with
signature (11,3) and how to recognize the 14 dimensions when the theory is
compactified to lower dimensions. \medskip

\section{From 10 to 14 dimensions.}

The $A(B)$ superalgebra in 10D contains two 16-component Majorana-Weyl
spinors of opposite (same) chirality $32_{A(B)}=16_L+16_{R(L)}.$ The $A$
superalgebra may be rewritten as an 11D superalgebra using a single 32
component spinor. In M-theory \cite{witten}, the anticommutator has all
possible 528$_A$ extensions corresponding to the 11D momentum and the 2 and
5 branes \cite{townsend} $528_A=11+55+462.$ This hides a 12D structure of
the form $528_A=66+462$ which corresponds to a 12D superalgebra \cite
{ibtokyo}-\cite{tseytlin}

\begin{equation}
\left\{ Q_\alpha ,Q_\beta \right\} =\gamma _{\alpha \beta
}^{M_1M_2}\,\,Z_{M_1M_2}\,+\gamma _{\alpha \beta }^{M_1\cdots
M_6}\,\,\,Z_{M_1\cdots M_6}^{+}  \label{typea}
\end{equation}
where $\alpha $ labels the 32 component Weyl spinor of SO(10,2). The (10,2)
signature is necessary to have a real spinor\footnote{
Bott periodicity indicates that the reality properties of spinors are
similar for SO$\left( 10,1\right) \sim $ SO$\left( 2,1\right) =$SL$\left(
2,R\right) $ and SO$\left( 10,2\right) \sim $ SO$\left( 2,2\right) =$SL$
\left( 2,R\right) \times \,$SL$\left( 2,R\right) $ and SO$\left( 11,2\right)
\sim $ SO$\left( 3,2\right) =$Sp$\left( 4,R\right) $ and SO$\left(
11,3\right) \sim $ SO$\left( 3,3\right) =$SL$\left( 4,R\right) $.}. The 12th
gamma matrix is the 32$\times 32$ identity matrix $\gamma ^{0^{\prime
}}=1_{32}$ since it is in the Weyl sector\footnote{
In terms of 11D one obtains $Z_{M_1M_2}\rightarrow P_\mu \oplus Z_{\mu _1\mu
_2}$ and $Z_{M_1\cdots M_6}^{+}\rightarrow Z_{\mu _1\cdots \mu _5}.$ If
non-zero, the 2 and 5 branes are sources coupled to the 11D antisymmetric
gauge potential $A_{\mu _0\mu _1\mu _2}$ and its magnetic dual $A_{\mu _0\mu
_1\mu _2\cdots \mu _5}$. The Lorentz generator $L_{\mu _1\mu _2}$ is outside
of this algebra since $Z_{\mu _1\mu _2}$ interpreted as above cannot
coincide with $L_{\mu _1\mu _2}.$ Therefore the 11 or 12D superalgebra is
not related to the superconformal algebra that has operators labelled in a
similar way. There is an extended superconformal algebra with the same
structure \cite{vanproyen} but it has different physical content. One may
wonder about the closure and Jacobi identities since such properties
determine the representations of the superalgebra. For our present
discussion this question may be left open since there are various
possibilities \cite{str96}. The simplest case is to take Abelian extensions
corresponding to a linearized flat limit of the theory. A model for a curved
space may correspond to OSp(1/32) which satisfies all Jacobi identities.
Intermediate cases are obtained by considering various contractions of
OSp(1/32). However, there are more possibilities since the operators may
close into an even larger set of operators in the quantum theory of
interacting p-branes. The anticommutator considered in our discussion
applies to all cases.}.

\medskip The 10D type-IIB superalgebra is not included in the 11D or 12D
superalgebra above. The $B$ superalgebra with its 528$_B$ bosonic generators
can be written in a form that exhibits higher dimensions \cite{stheory} 
\begin{equation}
\left\{ Q_{\bar{\alpha}\bar{a}},Q_{\bar{\beta}\bar{b}}\right\} =\left( i\tau
_2\tau _i\right) _{\bar{a}\bar{b}}\left[ \bar{\gamma}_{\bar{\alpha}\bar{\beta
}}^{\bar{\mu}}\,\,P_{\bar{\mu}}^i+\bar{\gamma}_{\bar{\alpha}\bar{\beta}}^{
\bar{\mu}_1\cdots \bar{\mu}_5}\,\,\,X_{\bar{\mu}_1\cdots \bar{\mu}
_5}^i\right] +\bar{\gamma}_{\bar{\alpha}\bar{\beta}}^{\bar{\mu}_1\bar{\mu}_2
\bar{\mu}_3}\,\,\left( i\tau _2\right) _{\bar{a}\bar{b}}\,Y_{\bar{\mu}_1\bar{
\mu}_2\bar{\mu}_3}.  \label{type2b}
\end{equation}
where $\bar{\alpha},\bar{\beta}=1,2,\cdots ,16$ and $\bar{a},\bar{b}=1,2$
are spinor indices, while $\bar{\mu}=0,1,\cdots ,9$ and $i=0^{\prime
},1^{\prime },2^{\prime }$ are vector indices for $SO(9,1)\times SO(2,1).$
The $\tau _i$ are given by Pauli matrices $\tau _i=\left( -i\tau _2,\tau
_1,\tau _3\right) .$ The $P_{\bar{\mu}}^i$ is a SO$(2,1)$ triplet containing
the momentum $p_{\bar{\mu}}$ and two 1-brane sources $w_{\bar{\mu}}^{1,2}$
that couple to the two antisymmetric gauge potentials $B_{\mu _0\mu
_1}^{1,2} $ in type-IIB superstring theory \footnote{
It may seem unusual that the 10D momentum operator is a member of the SO$
\left( 2,1\right) $ triplet $P_{\bar{\mu}}^i,$ but this is clearly true.
This SO$\left( 2,1\right) $ should not be confused with the U duality group
SL(2,Z). U acts on the gauge potentials $B_{\mu _0\mu _1}^{1,2}$ and sources 
$w_{\bar{\mu}}^{1,2}$ but leaves the metric and momentum $p_{\bar{\mu}}$
invariant. The transformation known as S duality fits in the maximal compact
subgroup of both the SO$\left( 2,1\right) $ and SL(2,Z). What is the
relation between these groups? To distentagle them one may expand $P_{\bar{
\mu}}^i=p_{\bar{\mu}}v^i+w_{\bar{\mu}}^1v_1^i+w_{\bar{\mu}}^1v_2^i$ by using
a basis of three orthogonal SO$\left( 2,1\right) \,\,$vectors $
(v^i,v_1^i,v_2^i)$ where $v^i$ is timelike and $v_{1,2}^i$ are spacelike.
Then $p_{\bar{\mu}}$ is the 10D momentum and $v^i$ is a 3D ``momentum'' \cite
{twotimes} both of which are singlet under SL(2,Z) while the spacelike
vectors ($w_{\bar{\mu}}^1,w_{\bar{\mu}}^2)$ and $(v_1^i,v_2^i)$ form
doublets under SL(2,Z) such that $P_{\bar{\mu}}^i$ is a singlet of SL(2,Z).
This is the $B$ basis that exhibits the Lorentz symmetry SO$\left(
2,1\right) $ while hiding the duality symmetry. In the the rest frame of the
internal ``momentum'' one may take $v^i=\left( 1,0,0\right) $ so that the
superalgebra takes the more familiar form $\left\{ Q_{\bar{\alpha}}^{\bar{a}
},Q_{\bar{\beta}}^{\bar{b}}\right\} =\bar{\gamma}_{\bar{\alpha}\bar{\beta}}^{
\bar{\mu}}\,\,\left( i\tau _2\tau _i\right) ^{\bar{a}\bar{b}}p_{\bar{\mu}
}v^i+\cdots =\bar{\gamma}_{\bar{\alpha}\bar{\beta}}^{\bar{\mu}}\,\,\delta ^{
\bar{a}\bar{b}}p_{\bar{\mu}}+\cdots $. This latter form exhibits the SL(2,Z)
basis rather than the Lorentz basis. Therefore the duality and the Lorentz
bases are related to each other by a boost in SO(2,1).
\par
This is a general situation that applies in every dimension as explained in 
\cite{sentropy}. Namely, the basis in which the momentum is a singlet under
dualities is defined to be the duality basis. In the $A$ or $B$ bases the
momentum is a member of a multiplet that transforms under an internal
Lorentz group. In the rest frame of an internal momentum one recovers the
duality basis. Hence $A$ and $B$ bases are connected to the duality basis
with boosts and they are related to each other by T-duality.}. In \cite
{stheory} the SO$(2,1)$ symmetry was interpreted as Lorentz transformations
in 3 additional dimensions beyond the usual 10D. In the present paper it
will be more properly interpreted as the SL$\left( 2,R\right) _{+}$ embedded
in SO$\left( 2,2\right) =$ SL$\left( 2,R\right) _{+}\times $ SL$\left(
2,R\right) _{-}.$ Then the triplet index $i$ will be replaced by the self
dual SO$\left( 2,2\right) $ tensor $\left[ mn\right] _{+}$ which is a
triplet of SL$\left( 2,R\right) _{+}$ when the vector index $m=11,12,13,14$
spans the extra 4 dimensions with signature $\left( +,-,+,-\right) .$

Based on T duality we suggest that the $A,B$ superalgebras are different
sectors of the same fundamental theory as explained in the previous section.
To unify them in the same theory one is led to 64 supercharges classified as
the spinor in 14D with signature $\left( 11,3\right) $ as a generalization
of S theory \cite{stheory}$.$ This is possible since SO$\left( 11,3\right) $
also has a Majorana-Weyl spinor of dimension 64 (see footnote 1).

There should exist a mechanism that provides two distinct projectors that
cuts down 64 to the branches 32$_{A,B}$. We will see soon that there is a
third distinct projector that leads to a third branch $C$ with 32$_C$
fermions. The heterotic and type-I superalgebras are secondary branches
attached to the main $A,B,C$ branches. As will be shown, all such branches
and sub-branches may be embedded {\it covariantly} in SO$\left( 11,3\right)
. $

We remind the reader of the non-covariant $A$ and $B$ projectors. By using
the SO$\left( 11,2\right) $ 64$\times $64 gamma matrices $\Gamma _M$ as in
the appendix of \cite{stheory}, the $A$ projector is given by $1+\Gamma
_{13}.$ It distinguishes the spacelike 13th dimension. The $B$ projector is (
$1+\Gamma _{11}\Gamma _{12}\Gamma _{13})=(1+\Gamma _0\cdots \Gamma _9).$ It
distinguishes the (2,1) from the (9,1) dimensions. Thus, as exhibited in the
superalgebras above, they are covariant under the groups $A=$SO$(10,2)$ and $
B=$SO$(9,1)\times $SO$(2,1)$ respectively, while they are both embedded in SO
$\left( 11,2\right) .$ How do these fit in 14D? It is possible to go up one
more dimension because the 64 component real spinor may also be regarded as
the Weyl spinor of SO$\left( 11,3\right) .$ In the Weyl sector the timelike
14th gamma matrix is given by the identity matrix $\Gamma _{14}=1_{64}.$
Then the projector to the $A$ sector is really lightlike ($\Gamma
_{14}+\Gamma _{13})$ leaving behind SO$\left( 10,2\right) $ covariance as
desired. From the point of view of 14D $\rightarrow \left( 9,1\right) +(2,2)$
note that the $A$ projector is a lightlike vector embedded in $\left(
2,2\right) .$ This projector cuts down 64 to 32$_A$ which consists of two
opposite chirality 16 component spinors in 10D.

For the $B$ sector consider also SO$\left( 11,3\right) \longrightarrow $ SO$
\left( 9,1\right) \times $ SO$\left( 2,2\right) $ and note that there are
two possibilities for embedding the SO$\left( 2,1\right) $ that appears in
eq.(\ref{type2b}). It could be identified either with the vectorial SL(2,R)$
_V\,\,$or with the chiral-like SL$(2,R)_{+}$ embedded in SO$\left(
2,2\right) $ 
\begin{equation}
{SO}\left( 2,2\right) ={SL}(2,R)_{+}\times {SL}
(2,R)_{-}\supset {SL}(2,R)_V.  \label{v}
\end{equation}
However, as we will see in footnote 5 below, there is no difference of
content between the two and the proper interpretation is the SL$(2,R)_{+}.$
This chiral embedding is fully covariant under SO$\left( 9,1\right) \times $
SO$\left( 2,2\right) $. \thinspace The $B$ projector which is consistent
with this invariance is ($1+\Gamma _{11}\Gamma _{12}\Gamma _{13}\Gamma
_{14})=(1+\Gamma _0\cdots \Gamma _9).$

Are there any other main branches? A main branch will be defined as a
superalgebra that has symmetries that cannot be contained as a subgroup of
the symmetries of another branch. The symmetries should be realized on only
32 {\it real} fermions, not 64. Only symmetries of the form SO$\left(
n,1\right) \times $ SO$\left( 11-n,2\right) $ or SO$\left( n,3\right) \times 
$ SO$\left( 11-n,0\right) $ that fit within SO$\left( 11,3\right) $ need to
be considered\footnote{
We have in mind a theory of various interacting p-branes formulated in some
way in terms of $X^M(\tau ,\sigma _1,\cdots ,\sigma _p),$ where the
coordinate index $M$ provides the basis for SO$\left( n,m\right) $
transformations. When we refer to ``dimensions'' we mean $X^M,$ and the
symmetries we are discussing are rotations of these coordinates.}. Since the
symmetries SO$\left( 10,2\right) $ and SO$\left( 9,1\right) \times $SO$
\left( 2,2\right) $ are not contained in each other, the $A,B$ branches are
distinct main branches. We have found that there is only one other main
branch with symmetry SO$\left( 3,3\right) \times $ SO$\left( 8\right) $ that
satisfies the $C$ superalgebra 
\begin{equation}
\left\{ Q_{\alpha a},Q_{\beta b}\right\} =\gamma _{\alpha \beta }^\mu \gamma
_{ab}^{ij}P_{\mu ij}+\gamma _{\alpha \beta }^{\mu _1\mu _2\mu _3}\delta
_{ab}T_{\mu _1\mu _2\mu _3}^{+}+\gamma _{\alpha \beta }^{\mu _1\mu _2\mu
_3}\gamma _{ab}^{ijkl}Z_{\mu _1\mu _2\mu _3ijkl}^{+}  \label{typec}
\end{equation}
Here $\alpha ,\beta $ are spinor indices for SO$\left( 3,3\right) =$ SL$
\left( 4,R\right) ,$ and $a,b$ are $8_{+}$ spinor indices for SO$\left(
8\right) $. Both of these spinors are Majorana-Weyl, therefore there are 4$
\times $8=32 real components. The indices $\mu ,i$ are vector indices for SO$
\left( 3,3\right) ,$ SO$\left( 8\right) $ respectively. The gamma matrices $
\gamma _{\alpha \beta }^{\mu _1\mu _2\mu _3},\gamma _{ab}^{ijkl}$ are
automatically self dual in the indices $[\mu _1\mu _2\mu _3]$ and [$ijkl]$
since they are in the Weyl sectors. Therefore the tensors $T_{\mu _1\mu
_2\mu _3}^{+},$ $Z_{\mu _1\mu _2\mu _3ijkl}^{+}$ are self dual in the
corresponding indices. The number of bosons may be counted as follows: for $
P_{\mu ij}$ 6$\times 28=168,$ for $T_{\mu _1\mu _2\mu _3}^{+}$ $\frac 12
\frac{6\times 5\times 4}{1\times 2\times 3}=10,$ for $Z_{\mu _1\mu _2\mu
_3ijkl}^{+}$ 10$\times \frac 12\frac{8\times 7\times 6\times 5}{1\times
2\times 3\times 4}=350.$ The total is 528$_C$.

The heterotic and type-I superalgebras in 10D may be obtained as
sub-branches of the main branches. Also different compactifications are
sub-branches. We will present them below as sub-branches in a 14D covariant
formalism of the main branches.

\section{14D covariance}

The superalgebra with 64 spinors may be written covariantly for SO$\left(
11,3\right) $ in the form $\left\{ Q_\alpha ,Q_\beta \right\} =S_{\alpha
\beta }^{64}\,\,\,$with 
\begin{equation}
S_{\alpha \beta }^{64}=\Gamma _{\alpha \beta }^{\bar{M}_1\bar{M}_2\bar{M}
_3}E_{\bar{M}_1\bar{M}_2\bar{M}_3}+\Gamma _{\alpha \beta }^{\bar{M}_1\bar{M}
_2\cdots \bar{M}_7}F_{\bar{M}_1\bar{M}_2\cdots \bar{M}_7}^{+},
\end{equation}
where $\bar{M}$ labels the vector of SO$\left( 11,3\right) .$ The tensor $
F^{+}$ is self dual because the seven index antisymmetric gamma matrix in
the Weyl sector is automatically self dual in 14D. This superalgebra
contains 2080 bosons (=364+1716). If reduced to 13D one obtains
antisymmetric tensors with 2,3,6 indices corresponding to the decomposition
2080 = 78+286+1716.

We suggest that, independent of the details of the fundamental theory, the
mechanism that cuts 64$\rightarrow 32$ can be formulated as a SO$\left(
11,3\right) $ covariant BPS equation in the form 
\begin{equation}
\det \left( S_{\alpha \beta }^{64}\right) =0.  \label{bps}
\end{equation}
The multiplicity of the zero eigenvalues of this equation corresponds to the
number of supercharges that vanish on the BPS states.

\subsection{A branch}

The solution of eq.(\ref{bps}) corresponding to the $A$ branch may be
written in a 14D covariant notation as follows 
\begin{eqnarray}
\left( S_A^{64}\right) _{\alpha \beta } &=&\Gamma _{\alpha \beta }^{\bar{M}_1
\bar{M}_2\bar{M}_3}Z_{\bar{M}_1\bar{M}_2}V_{\bar{M}_3}+\Gamma _{\alpha \beta
}^{\bar{M}_1\bar{M}_2\cdots \bar{M}_7}Z_{\bar{M}_1\bar{M}_2\cdots \bar{M}
_6}^{+}V_{\bar{M}_7}\,, \\
V_{\bar{M}}V^{\bar{M}} &=&Z_{\bar{M}_1\bar{M}_2}V^{\bar{M}_2}=Z_{\bar{M}_1
\bar{M}_2\cdots \bar{M}_6}^{+}V^{\bar{M}_6}=0,  \nonumber
\end{eqnarray}
where $V_{\bar{M}}$ is a 14D lightlike vector and the antisymmetric tensors $
Z,Z^{+}$ are orthogonal to it. Thanks to orthogonality one may factor out
the lightlike projector $\Gamma _{\bar{M}}V^{\bar{M}}$ as an overall factor
in $S_A^{64}$, showing that 32 supercharges vanish$.$ Although orthogonality
allows components in $Z,Z^{+}$ that point along $V_{\bar{M}},$ those drop
out due to the antisymmetric indices $\bar{M}_k$ on $\Gamma _{\alpha \beta
}^{\bar{M}_1\bar{M}_2\bar{M}_3},\,\,\Gamma _{\alpha \beta }^{\bar{M}_1\bar{M}
_2\cdots \bar{M}_7}$. The remaining effective subspace of $\left(
10,2\right) $ indices orthogonal to $V_{\bar{M}}$ give precisely 528$_A$
bosons. $Z_{\bar{M}_1\bar{M}_2\cdots \bar{M}_6}^{+}$ is self dual in the $
\left( 10,2\right) $ subspace because the six-index gamma matrices in the
14D Weyl basis $\Gamma _{\alpha \beta }^{\bar{M}_1\bar{M}_2\cdots \bar{M}
_7}V_{\bar{M}_7}$ are automatically self dual in the $\left( 10,2\right) $
subspace thanks to the lightlike vector.

Using the SO$\left( 11,3\right) $ covariance one can choose a special frame
in which $V_{\bar{M}}$ points along the 13th+14th dimensions with signature $
\left( 1,1\right) $. Then $\Gamma _{\bar{M}}V^{\bar{M}}\sim \Gamma
_{13}+\Gamma _{14}$ and $S_A^{64}$ takes the block diagonal form $
S_A^{64}=\left( S_A^{32},0\right) $ where we have used $\Gamma ^{14}=1_{64}.$
Then $S_A^{32}$ is precisely the right hand side of eq.(\ref{typea}).

In a general frame the full SO$\left( 11,3\right) $ covariance is possible
provided the lightlike vector is an operator rather than a fixed vector
frozen in some direction. The presence of such operators lead to more
general superalgebras. Examples, possible interpretations and physical roles
of such operators may be found in \cite{superpv} \cite{twotimes}.

\subsection{B branch}

Introduce four orthogonal SO$\left( 11,3\right) $ unit vectors $V_{\bar{M}
}^m,$ with $m=1,2,3,4.$ Let $V^1,V^3$ be spacelike and $V^2,V^4$ timelike.
Then the $m$ label defines an auxiliary SO$\left( 2,2\right) ^{\prime }.$
Construct SO$\left( 11,3\right) \times $ SO$\left( 2,2\right) ^{\prime }$
covariant antisymmetric tensors 
\begin{equation}
F_{\bar{M}\bar{N}}^{\left[ mn\right] _{+}}=V_{[\bar{M}}^mV_{\bar{N}
]}^n+\frac 12\varepsilon _{\,\,\,\,\,\,\,\,\,pq}^{mn}V_{[\bar{M}}^pV_{\bar{N}
]}^q.
\end{equation}
They are {\it self dual} tensors for the auxiliary SO$\left( 2,2\right)
^{\prime }=$SL$\left( 2,R\right) _{+}^{\prime }\times $ SL$\left( 2,R\right)
_{-}^{\prime }$, so they form a {\it triplet} of the auxiliary SL$\left(
2,R\right) _{+}^{\prime }$. The covariant superalgebra for the $B$ sector
can now be written by giving the solution to the BPS equation in the form 
\begin{eqnarray}
\left( S_B^{64}\right) _{\alpha \beta } &=&\Gamma _{\alpha \beta }^{\bar{M}_1
\bar{M}_2\cdots \bar{M}_7}\left( Z_{\bar{M}_1\bar{M}_2\cdots \bar{M}
_5}^{\left[ mn\right] _{+}}F_{\bar{M}_6\bar{M}_7\left[ mn\right] _{+}}\,+Y_{
\bar{M}_1\bar{M}_2\bar{M}_3}\,V_{\bar{M}_4}^1V_{\bar{M}_5}^2V_{\bar{M}
_6}^3V_{\bar{M}_7}^4\right)  \nonumber \\
&&+\Gamma _{\alpha \beta }^{\bar{M}_1\bar{M}_2\bar{M}_3}\left( P_{\bar{M}
_1}^{\left[ mn\right] _{+}}F_{\bar{M}_2\bar{M}_3\left[ mn\right] _{+}}+Y_{
\bar{M}_1\bar{M}_2\bar{M}_3}\right) ,
\end{eqnarray}
where the bosons $P,Y,Z$ are orthogonal to all four vectors 
\begin{equation}
P_{\bar{M}_1}^{\left[ pq\right] _{+}}V_m^{\bar{M}_1}=Y_{\bar{M}_1\bar{M}_2
\bar{M}_3}V_m^{\bar{M}_3}=Z_{\bar{M}_1\bar{M}_2\cdots \bar{M}_5}^{\left[
pq\right] _{+}}V_m^{\bar{M}_5}=0.
\end{equation}
Due to orthogonality, the $Y_{\bar{M}_1\bar{M}_2\bar{M}_3}$ term in $
S_B^{64} $ is proportional to the projector 
\begin{equation}
\left( 1+V^1V^2V^3V^4\right) ,
\end{equation}
where the $V^m\equiv V_{\bar{M}}^m\Gamma ^{\bar{M}}$ anticommute $
V^1V^2=-V^2V^1.$ Also, because of the self duality of $F_{\bar{M}_6\bar{M}
_7\left[ mn\right] _{+}}$ the other terms may be multiplied by the same
projector without changing anything. Therefore, the full $S_B^{64}$ is
proportional to the same projector, signaling the vanishing of 32
supercharges.

The projector is invariant under SO$\left( 11,3\right) \times $ SO$\left(
2,2\right) ^{\prime }$. This symmetry can be used to gauge fix the $V_{\bar{M
}}^m$ to a special frame $V_{\bar{M}}^m\rightarrow \delta _{\bar{M}}^{10+m}$
in which the orthogonal 14D vectors $V_{\bar{M}}^1,V_{\bar{M}}^2,V_{\bar{M}
}^3,V_{\bar{M}}^4$ point along the 11th,12th,13th and 14th dimensions
respectively. In this frame the auxiliary SO$\left( 2,2\right) ^{\prime }$
coincides with the SO$\left( 2,2\right) $ embedded in SO$\left( 11,3\right) $
. The projector becomes $\left( 1+\Gamma _{11}\Gamma _{12}\Gamma _{13}\Gamma
_{14}\right) $ = $\left( 1+\Gamma _0\cdots \Gamma _9\right) $, which is
recognized as the $B$ projector of the previous section. The superalgebra
then collapses to the non-covariant form in eq.(\ref{type2b}) with remaining
explicit symmetry SO$\left( 9,1\right) \times $ SO$\left( 2,2\right) ,$
after replacing the triplet index $i$ with the triplet index $\left[
mn\right] _{+}$.

In the general frame there are 32$_B$ fermions and 528$_B$ bosons $P,Y,Z$
covariantly embedded in SO$\left( 11,3\right) .$ In addition there are also
the 56(=4$\times $14) components of the vectors $V_{\bar{M}}^m.$ For the
full covariance to be valid these must be operators rather than fixed
vectors, as in the examples of \cite{superpv} \cite{twotimes}.

\subsection{C branch}

Consider 8 orthogonal spacelike unit vectors $V_{\bar{M}}^i.$ There is an
auxiliary SO$\left( 8\right) ^{\prime }$ defined on the indices $i.$ The $C$
branch solution to the BPS equation is covariant under SO$\left( 11,3\right)
\times $ SO$\left( 8\right) ^{\prime }$ 
\begin{eqnarray}
\left( S_C^{64}\right) _{\alpha \beta } &=&\Gamma _{\alpha \beta }^{\bar{M}_1
\bar{M}_2\bar{M}_3}\left( P_{\bar{M}_1ij}V_{\bar{M}_2}^iV_{\bar{M}_3}^j+T_{
\bar{M}_1\bar{M}_2\bar{M}_3}^{+}\right) \\
&&+\Gamma _{\alpha \beta }^{\bar{M}_1\bar{M}_2\cdots \bar{M}_7}\,\,Z_{\bar{M}
_1\bar{M}_2\bar{M}_3ijkl}^{+}V_{\bar{M}_2}^iV_{\bar{M}_3}^jV_{\bar{M}_2}^kV_{
\bar{M}_3}^l  \nonumber \\
&&+\frac 1{7!}\Gamma _{\alpha \beta }^{\bar{M}_1\bar{M}_2\cdots \bar{M}_7}P_{
\bar{M}_1}^{i_8i_1}\,\,V_{\bar{M}_2}^{i_2}\cdots V_{\bar{M}
_7}^{i_7}\varepsilon _{i_8i_1\cdots i_7}
\end{eqnarray}
where the tensors $P,T,Z$ are orthogonal to the $V_{\bar{M}}^i.$ There are
self duality conditions: $Z_{\bar{M}_1\bar{M}_2\bar{M}_3ijkl}^{+}$ is SO$
\left( 8\right) ^{\prime }$ self dual in the [$ijkl$] indices, and the
tensor $T_{\bar{M}_1\bar{M}_2\bar{M}_3}^{+}$ satisfies the 14D duality
condition 
\begin{equation}
T^{+\bar{M}_1\bar{M}_2\bar{M}_3}=\frac 1{3!}\varepsilon ^{\bar{M}_1\bar{M}_2
\bar{M}_3\bar{N}_1\bar{N}_2\bar{N}_3K_1\cdots K_8}T_{\bar{N}_1\bar{N}_2\bar{N
}_3}^{+}V_{\bar{K}_1}^1\cdots V_{\bar{K}_8}^8.
\end{equation}
Due to orthogonality the terms involving $P$ can be rewritten in the form 
\begin{equation}
\Gamma _{\alpha \beta }^{\bar{M}_1\bar{M}_2\bar{M}_3}P_{\bar{M}_1ij}V_{\bar{M
}_2}^iV_{\bar{M}_3}^j\left( 1+V^1\cdots V^8\right) .
\end{equation}
where the $V^i\equiv V_{\bar{M}}^i\Gamma ^{\bar{M}}$ anticommute among
themselves $V^1V^2=-V^2V^1.$ Using the duality properties of $T,Z$ given
above, the remaining terms in $S_C^{64}$ may be multiplied by the same
projector without changing anything. Therefore the full $S_C^{64}$ is
proportional to the same projector, showing that only 32$_C$ supercharges
survive. If one recalls that $\Gamma _{\alpha \beta }^{\bar{M}_1\bar{M}
_2\cdots \bar{M}_7}$ is self dual in 14D then the presence of the projector
forces $Z_{\bar{M}_1\bar{M}_2\bar{M}_3ijkl}^{+}$ to satisfy a duality
condition on the $\bar{M}$ indices similar to the one satisfied by $T^{+\bar{
M}_1\bar{M}_2\bar{M}_3}.$ One may then verify that the number of bosons $
P,T,Z$ is $528_C.$ In addition there are 112 (=8$\times $14) bosons
describing the spacelike unit vectors $V_{\bar{M}}^i.$

The projector $\left( 1+V^1\cdots V^8\right) $ is invariant under SO$\left(
11,3\right) \times $ SO$\left( 8\right) ^{\prime }.$ Using the symmetry one
may gauge fix to $V_{\bar{M}}^i=\delta _{\bar{M}}^i.$ The projector becomes $
\left( 1+\Gamma ^1\cdots \Gamma ^8\right) =\left( 1+\Gamma ^9\Gamma ^0\Gamma
^{11}\Gamma ^{12}\Gamma ^{13}\Gamma ^{14}\right) $ showing that it projects
to the self dual sectors of SO$\left( 6,6\right) \times $ SO$\left( 8\right) 
$ which is the surviving symmetry in the special frame. Then the
superalgebra $C$ collapses to the non-covariant form of eq.(\ref{typec}).

\subsection{Heterotic and type-I sub-branches}

Just as the $32_{A,B,C}$ fermions are distinct embeddings in 64, the 528$
_{A,B,C}$ bosons are distinct embeddings in the 2080. However, some of the
fermions and bosons are common among the various sets. To identify the
common ones one may use the following method. Consider the $A,B$ subsets in
the special frames and use the SO$\left( 9,1\right) \times $ SL$\left(
2,R\right) _{+}\times $ SL$\left( 2,R\right) _{-}$ basis embedded in SO$
\left( 11,3\right) $ \footnote{
Due to the lack of space we will not discuss the $CB$ or $CA$ pairs. The
method is similar and a useful common basis is SO$\left( 8\right) \times $ SO
$\left( 1,1\right) \times $ SO$\left( 2,2\right) .$}. The 64 Weyl fermions
and 2080(=364+1716) bosons are reclassified as follows 
\begin{eqnarray}
64 &=&\left( 16_{+},2,1\right) +\left( 16_{-},1,2\right) ,  \nonumber \\
364 &=&\left( 120,1,1\right) +\left( 45,2,2\right) +\left( 10,3,1\right)
+\left( 10,1,3\right) +\left( 1,2,2\right) , \\
1716 &=&\left( 120,1,1\right) ^{\prime }+\left( 210,2,2\right) +\left(
126_{+},3,1\right) +\left( 126_{-},1,3\right) .  \nonumber
\end{eqnarray}

The $A$ projection keeps two 10D fermions of opposite chirality $32_A=\left(
16_{+},+,1\right) +\left( 16_{-},1,+\right) $ where we have denoted each $
2=\pm $ and kept only the $+$ component. The $A$ projection gives the bosons 
\begin{eqnarray}
528_A &=&\left( 45,+,+\right) +\left( 10,++,1\right) +\left( 10,1,++\right)
+\left( 1,+,+\right) \\
&&+\left( 210,+,+\right) +\left( 126_{+},++,1\right) +\left(
126_{-},1,++\right)  \nonumber
\end{eqnarray}

The $B$ projector keeps two 10D fermions of the same chirality which also
happen to be singlets of SL$\left( 2,R\right) _{-},$ namely 32$_B=\left(
16_{+},2,1\right) .$ Then the 528$_B$ bosons are\footnote{
Can one find a $B$ projector that identifies the SO$\left( 2,1\right) $ of
eq.(\ref{type2b}) with SL$\left( 2,R\right) _V$ of eq.(\ref{v}) rather than
SL$\left( 2,R\right) _{+}?$ To answer the question reduce the
representations of SL$\left( 2,R\right) _{+}\times $ SL$\left( 2,R\right)
_{-}$ to SL$\left( 2,R\right) _V$ and then pick the the two $16_{+}$
fermions. Evidently the fermions are the same ones since there are only two
of them. But then the bosons must also be unambigously the same as those in
eq.(\ref{528b}) since they appear in the products of the same fermions that
were previously classified as $\left( 16_{+},2,1\right) .$ Therefore there
is no other $B$ projector, and the symmetry of eq.(\ref{type2b}) is
automatically SL$\left( 2,R\right) _{+}$ embedded in SO$\left( 2,2\right) ,$
not SL$\left( 2,R\right) _V.$} 
\begin{equation}
528_B=\frac 12\left[ \left( 120,1,1\right) +\left( 120,1,1\right) ^{\prime
}\right] +\left( 10,3,1\right) +\left( 126_{+},3,1\right) .  \label{528b}
\end{equation}

By comparing the two sets 528$_{A,B}$ one finds that the common subset is
the heterotic fermions $\left( 16_{+},+,1\right) $ and bosons $\left(
10,++,1\right) +\left( 126_{+},++,0\right) $ which form the heterotic
superalgebra in 10D. This provides the key for the embedding of the
heterotic superalgebra covariantly in 14D. It is obtained by starting with
the solution $S_B^{64}$ and keeping only the bosons that couple to the
combination $F_{\bar{M}\bar{N}}^{+}\equiv \frac 12\left( F_{\bar{M}\bar{N}
}^{42}+F_{\bar{M}\bar{N}}^{41}\right) $ while setting all others equal to
zero. This combination picks up only the components $\left( 10,++,1\right)
+\left( 126_{+},++,1\right) $ among the bosons when written in the special
frame. More simply, one can show that $F_{\bar{M}\bar{N}}^{+}=V_{[\bar{M}}V_{
\bar{N}]}^{\prime }$ where $V_{\bar{M}}\sim \frac 1{\sqrt{2}}\left( V_{\bar{M
}}^4+V_{\bar{M}}^3\right) $ and $V_{\bar{M}}\sim \frac 1{\sqrt{2}}\left( V_{
\bar{M}}^2+V_{\bar{M}}^1\right) $ are the lightcone combinations which are
orthogonal but not parallel to each other $V\cdot V=V^{\prime }\cdot
V^{\prime }=V\cdot V^{\prime }=0.$ In terms of these the heterotic
superalgebra is embedded covariantly in 14D as follows 
\begin{eqnarray}
\left( S_{Het}^{64}\right) _{\alpha \beta } &=&\Gamma _{\alpha \beta }^{\bar{
M}_1\bar{M}_2\bar{M}_3}P_{\bar{M}_1}V_{\bar{M}_2}V_{\bar{M}_3}^{\prime
}+\Gamma _{\alpha \beta }^{\bar{M}_1\bar{M}_2\cdots \bar{M}_7}\,\,Z_{\bar{M}
_1\cdots \bar{M}_5}^{+}V_{\bar{M}_6}V_{\bar{M}_7}^{\prime }, \\
P_{\bar{M}_1}V^{\bar{M}_1} &=&P_{\bar{M}_1}V^{\prime \bar{M}_1}=\,Z_{\bar{M}
_1\cdots \bar{M}_5}^{+}V^{\bar{M}_5}=Z_{\bar{M}_1\cdots \bar{M}
_5}^{+}V^{\prime \bar{M}_5}=0.  \nonumber
\end{eqnarray}
Although orthogonality permits components in $P,Z^{+}$ along the lightlike
directions $V,V^{\prime },$ these drop out due to the antisymmetry of the
gamma matrices. There remains an effective $\left( 9,1\right) $ subspace. $
Z^{+}$ is automatically self dual in the $\left( 9,1\right) $ subspace
thanks to the lightlike nature of the vectors in $\Gamma _{\alpha \beta }^{
\bar{M}_1\bar{M}_2\cdots \bar{M}_7}\,\,V_{\bar{M}_6}V_{\bar{M}_7}^{\prime }$
and the self duality of $\Gamma _{\alpha \beta }^{\bar{M}_1\bar{M}_2\cdots 
\bar{M}_7}$ in the 14D Weyl sector. The number of independent components in $
P,Z^{+}$ is 136.

Using orthogonality one sees that $S_{Het}^{64}$ is proportional to the
projector $\left( V\cdot \Gamma \right) \left( V^{\prime }\cdot \Gamma
\right) $ which is invariant under SO$\left( 11,3\right) .$ The double
lightcone projections cut down 64 to 16 non-zero supercharges. In the
special frame the projector becomes $\left( \Gamma ^{12}+\Gamma ^{11}\right)
\left( \Gamma ^{14}+\Gamma ^{13}\right) $ showing that the remaining
symmetry is SO$\left( 9,1\right) $ and that the superalgebra reduces to the
usual heterotic superalgebra in the non-zero 16$\times $16 block embedded in
64$\times 64.$

The type-I superalgebra may be obtained from the $B$ superalgebra by
identifying the two supercharges $\left( 16_{+},2,1\right) $. In the
non-covariant eq.(\ref{type2b}) this requires a right hand side that is
proportional to $\delta _{\bar{a}\bar{b}},$ which means only the $
i=0^{\prime }$ term contributes since $\left( \tau _2\tau _2\right) _{\bar{a}
\bar{b}}=\delta _{\bar{a}\bar{b}}.$ In the SO$\left( 2,2\right) $ notation
this is equivalent to the term $\left[ mn\right] _{+}=\left[ 42\right]
_{+}=\left[ 13\right] _{+}.$ Thus, the covariant embedding of the type-I
superalgebra in 14D is 
\begin{equation}
\left( S_I^{64}\right) _{\alpha \beta }=\Gamma _{\alpha \beta }^{\bar{M}_1
\bar{M}_2\bar{M}_3}P_{\bar{M}_1}F_{\bar{M}_2\bar{M}_3}^{42}+\Gamma _{\alpha
\beta }^{\bar{M}_1\bar{M}_2\cdots \bar{M}_7}\,\,Z_{\bar{M}_1\cdots \bar{M}
_5}^{+}F_{\bar{M}_6\bar{M}_7}^{42}
\end{equation}
Due to orthogonality there is an overall $F_{\bar{M}\bar{N}}^{42}\Gamma ^{
\bar{M}\bar{N}}$ factor that corresponds to the projector. In the special
frame the projector becomes $F_{\bar{M}\bar{N}}^{42}\Gamma ^{\bar{M}\bar{N}
}=\left( \Gamma ^{14}\Gamma ^{12}+\Gamma ^{13}\Gamma ^{11}\right) .$ This
seems to cut down 64 to 32, not to 16. However, the non-zero 32$\times 32$
sector is equivalent to two identical 16$\times 16$ blocks, showing that
only 16 supercharges of type-I are non-zero. The two identical block
structure is produced because the two 16$_{+}$'s were identified.

\subsection{Compactifications}

Every branch has compactifications down to 4D with SO$\left( 3,1\right) $
Lorentz symmetry and the internal symmetries inherited from the $A,B,C$ main
branches (see e.g. \cite{superpv} where the compactified $A,B$ branches in
4D are written out explicitly). Therefore one finds that the same 4D
extended superalgebra has several reclassifications which might be called $
A,B,C,$ each one containing a maximum of 32 fermions and 528 bosons,
corresponding to a maximum of 8 supersymmetries in 4D $.$ Each one has also
a duality basis which is obtained by a boost transformation in the internal
Lorenz dimensions (see \cite{sentropy} and footnote 3 ). The maps among
these classifications are related to various duality transformations. The
compactification sub-branches to various dimensions will not be discussed
here since they are obtained by naively compactifying the main branches.

We have shown that all known superalgebras derive from a single one in 14D
as solutions to the BPS equation. To exhibit the 14D nature of the solutions
some vectors $V$ were needed. If the various vectors $V$ are frozen
constants the symmetry is broken to some subgroup of SO$\left( 11,3\right) ,$
and the superalgebra collapses to the familiar one. If the vectors $V$ are
also operators then the SO$\left( 11,3\right) $ symmetry is not broken, on
the contrary there are more symmetries that we called ``auxiliary''.
Possible interpretations of such operators have been given elsewhere in
several contexts \cite{sentropy} \cite{superpv} \cite{twotimes}.

Our construction leads us to speculate that a fundamental theory in $\left(
11,3\right) $ dimensions may be behind string and p-brane duality
properties. The presence of extra timelike dimensions seems to pose a
problem for such a theory. However, in recent papers \cite{twotimes} we have
shown that extra timelike dimensions can be interpreted without the obvious
problems. The interpretation is done in the context of models that involve
several interacting particles or p-branes forming different physical
sectors, each with its own timelike dimension. The superalgebra of such
systems has the types of signatures and operators $V$ discussed in this
paper. The models have sufficient gauge symmetries to eliminate redundant
timelike degrees of freedom. A cosmological scenario may also be invoked to
arrive at the single time sector which describes our current universe.

Note Added: While this paper was under preparation ref.\cite{sezgin2}
appeared. This work was apparently stimulated by a brief announcement in 
\cite{twotimes} of our present work on $\left( 11,3\right) $ dimensions. The
embedding of 10D Yang-Mills theory in $\left( 11,3\right) $ dimensions lends
support to the ideas expressed here.

\end{document}